\documentclass[11pt]{article}
\usepackage{verbatim,color,amssymb,epsfig,epsf,graphicx,indentfirst,caption,lscape,multirow}
\usepackage{amscd,amsmath,amstext,amsfonts,amsbsy,amsthm}
\usepackage[width=.9\textwidth, font = small, labelfont = bf, labelsep = endash, textfont = it]{caption}

\begin{document}
\begin{center}
{\Large{\bf Improving likelihood-based inference in control rate regression}}\\
{Annamaria Guolo}
\\{University of Padova\\
Via Cesare Battisti 241/243, I-35121, Padova, Italy}
\\{annamaria.guolo@unipd.it}
\end{center}

\begin{center}
ABSTRACT 
\end{center}
Control rate regression is a diffuse approach to account for heterogeneity among studies in meta-analysis by including information about the outcome risk of patients in the control condition. Correcting for the presence of measurement error affecting risk information in the treated and in the control group has been recognized as a necessary step to derive reliable inferential conclusions. Within this framework, the paper considers the problem of small sample size as an additional source of misleading inference about the slope of the control rate regression. Likelihood procedures relying on first-order approximations are shown to be substantially inaccurate, especially when dealing with increasing heterogeneity and correlated measurement errors. We suggest to address the problem by relying on higher-order asymptotics. In particular, we derive Skovgaard's statistic as an instrument to improve the accuracy of the approximation of the signed profile log-likelihood ratio statistic to the standard normal distribution. The proposal is shown to provide much more accurate results than standard likelihood solutions, with no appreciable computational effort. The advantages of Skovgaard's statistic in control rate regression are shown in a series of simulation experiments and illustrated in a real data example. \texttt R code for applying first- and second-order statistic for inference on the slope on the control rate regression is provided.

\section{Introduction}
Control rate regression is a diffuse instrument in meta-analysis of clinical trials comparing a treated group and a control group (\cite{schmid1998}, \cite{arends}, \cite{vanhouwelingen}, \cite{chaimani}) to account for the between-study heterogeneity due to study designs, patients' characteristics and treatment interventions. In control rate regression, meta-analysis is performed by including 
a measure of the outcome risk of patients in the control condition, so that emerging differences among studies are a consequence of treatment effects only. The control rate, i.e., the proportion of patients with the event of interest in the control condition, represents a surrogate for the true risk of patients in the control condition. As a consequence, control rate is a measure affected by error. Properly accounting and correcting for the presence of measurement error is a necessary step to guarantee reliable inferential procedures (\cite{carroll}, \cite{buonaccorsi}) and avoid consequences such as biased estimates and inaccurate coverage probabilities of confidence intervals.

In this paper, we focus on likelihood-based procedures for measurement error correction in control rate regression. Advantages of the likelihood approach, mainly related to its limit properties, have been highlighted in Arends et al. \cite{arends}, Ghidey et al. \cite{ghidey2007}, and Guolo \cite{guolo2013}. In this paper we show that, despite the advantages in terms of properties of the maximum likelihood estimator, the likelihood approach suffers from several drawbacks. When relying on first-order approximations, such as, for example, the $\chi^2$ distribution for the likelihood ratio statistic, results can be seriously inaccurate in case of small sample size (e.g., \cite{brazzale}), that is, when the number of studies included in the meta-analysis is small. We suggest to overcome the problem and refine first-order likelihood inference through Skovgaard's second-order statistic \cite{skovgaard}. The present work takes advantage of previous results illustrated in Guolo \cite{guolo2012} within the classical meta-analysis framework and constitutes a step forward for developing Skovgaard's second-order statistic in the multivariate meta-analysis accounting for measurement errors. 
The accuracy of the results is obtained with no substantial computational effort, as the computation of Skovgaard's statistic components has a complexity comparable to that of evaluating the expected information matrix. Advantages over first-order results are highlighted in a series of simulation studies. The application of the method is illustrated via a real data example about the efficacy of a drug treatment against cardiovascular mortality in middle-aged patients with mild to moderate hypertension. 

The paper is structured as follows. Section~\ref{sec:framework} describes likelihood inference in control rate regression, by distinguishing first-order statistic and second-order statistic. Section~\ref{sec:simulation} illustrates the simulation studies used to evaluate the performance of the competing methods, while real data analysis is described in Section~\ref{sec:example}. The paper ends with some remarks in Section~\ref{sec:conclusion}. Technical details and additional simulation results are included in the Supplementary Material. The \texttt R \cite{r} code for implementing Skovgaard's second-order statistic is made available and illustrated in the Supplementary Material.

\section{Control rate regression}\label{sec:framework}
We consider a meta-analysis of $n$ independent studies about the effectiveness of a treatment. Let $\eta_i$ denote the risk measure in the treated group, or the treatment effect, and let $\xi_i$ denote the underlying risk measure in the control group, $i = 1, \ldots, n$. Control rate regression is typically a linear regression model (e.g., \cite{schmid1998}, \cite{arends})
\begin{equation}\label{eqn:model1}
\eta_i = \beta_0 + \beta_1 \xi_i + \varepsilon_i, \ \varepsilon_i \sim N(0, \tau^2),
\end{equation}
with parameter $\tau^2$ accounting for the heterogeneity with respect to the treatment measure in the population with the same underlying risk. The inferential interest is usually in $\beta_1$, with $\beta_1=0$ used to verify the constance of the treatment effect and its independence with respect to $\xi_i$.
An alternative specification of the model considers the relationship between the treatment effect $\eta_i-\xi_i$ and $\xi_i$ (e.g., \cite{ghidey2013}), with $(\beta_0,\beta_1)^\top = (0,1)^\top$ representing a claim of no relationship between the treatment effect and the risk in the control condition, on average.

The simplest approach for analysis suggested by Brand and Kragt \cite{brand} is a weighted least squares regression, with weights given by the inverse of the variance of the treatment effect. This approach does not consider that the summary information from each study represents a surrogate for the true unobserved risk measure and consequently is prone to measurement error. A huge literature
focuses on measurement error consequences, see Carroll et al. \cite{carroll} and Buonaccorsi \cite{buonaccorsi}. It has long been recognized that misleading inferential conclusions due to ignoring measurement errors include biased estimators, reduced power of tests, and inaccurate coverage probabilities of confidence intervals.
\\ 
Let $\hat \eta_i$ and $\hat \xi_i$ denote the observed error-prone versions of $\eta_i$ and $\xi_i$ available from study $i$. A commonly adopted measurement error structure (\cite{schmid1998}, \cite{arends}, \cite{schmid2004}) relates $(\hat\eta_i,\hat\xi_i)^\top$ to $(\eta_i,\xi_i)^\top$ through the bivariate normal distribution
\begin{equation}\label{eqn:mem}
\left(\begin{array}{c}
\hat\eta_i \\
\hat\xi_i
\end{array}\right)
\sim N_2 \left( 
\left(
\begin{array}{c}
\eta_i \\
\xi_i
\end{array}\right),
\Gamma_i
\right),
\end{equation}
where the within-study variance/covariance matrix $\Gamma_i$ is assumed to be known and estimated within each single study. Together with the specification of the regression model (\ref{eqn:model1}) and the measurement error model, the likelihood-based approach for inference requires the specification of the distribution for the underlying risk $\xi_i$. From a computational point of view, the most convenient choice is a normal model, $\xi_i \sim N(\mu, \sigma^2)$ (e.g., \cite{vanhouwelingen}).
Given the above distributional assumptions, the likelihood function for the whole parameter vector $\theta=(\beta_0, \beta_1, \mu,\tau^2, \sigma^2)^\top$ is obtained with a closed-form considering that, marginally, 

\begin{equation}\label{eqn:model}
\left(\begin{array}{c}
\hat\eta_i \\
\hat\xi_i
\end{array}\right)
\sim N_2 \left( 
\left(\begin{array}{c}
\beta_0+\beta_1 \mu \\
\mu
\end{array}\right),
\Gamma_i + \left( 
\begin{array}{cc}
\tau^2+\beta_1^2\sigma^2 & \beta_1 \sigma^2 \\
\beta_1 \sigma^2 & \sigma^2 
\end{array}
\right)
\right) .
\end{equation}
The computational convenience of the closed-form for the likelihood function is a practical justification for the choice of the normal specification for the measurement error model and for the underlying risk distribution. Different structures for both the models have been examined in the literature. Specification (\ref{eqn:mem}) is often an approximation of the exact measurement error structure, which can be defined case by case \cite{arends}, although at the price of computational complications. See also \cite{stijnen} for a detailed treatment of approximate and exact models in random-effects meta-analysis.
Alternatives to the normal model for the underlying risk include flexible solutions based on mixture of normals \cite{arends}, semiparametric specification \cite{ghidey2007} and the skew-normal distribution \cite{guolo2013}.

\section{First-order and higher-order likelihood inference}\label{sec:framework}
Consider the parameter vector $\theta=(\beta_0, \beta_1, \mu,\tau^2, \sigma^2)^\top$ introduced in the previous section. For convenience purposes, $\theta$ can be partitioned into a scalar component of interest $\psi$ and a remaining nuisance component $\lambda$, so that $\theta=(\psi, \lambda)^\top$. In control rate regression, starting from (\ref{eqn:model}), typically the inferential interest is on the parameter $\beta_1$ relating the treatment effect and the underlying risk measure. In this way, $\psi=\beta_1$ and $\lambda=(\beta_0, \mu, \tau^2, \sigma^2)^\top$. Let $\hat \theta=(\hat \psi, \hat \lambda)^\top$ denote the maximum likelihood estimate of $\theta$ and let $\tilde \theta=(\psi, \hat \lambda_{\psi})^\top$ denote the constrained maximum likelihood estimate of $\theta$ obtained for fixed $\psi$. \\
When inference is on a scalar component, then procedures can rely on the profile log-likelihood function $\ell_P(\psi)=\ell(\psi; \lambda_{\psi})$. Hypothesis testing and construction of confidence intervals can be based on the signed (square root of the) profile log-likelihood ratio statistic  
\begin{equation}\label{eqn:r}
r_P(\psi)= \rm{sign}\left(\hat\psi- \psi\right)\sqrt{2\left\{\ell_P(\hat\psi)-\ell_P(\psi)\right\}},
\end{equation}
which is preferable to the commonly adopted Wald-type statistic as the inferential procedures are invariant to reparameterization and confidence intervals based on $r_P$ are not forced to be symmetric.
Under mild regularity conditions, $r_P$ has an approximate standard normal distribution up to first-order error, see Section 4.4 in Severini \cite{severini}. In this way, a first-order accuracy $(1 - \alpha)\%$ confidence interval for $\psi$ is given by all the values satisfying $z_{\alpha/2} < r_P(\psi) < z_{1-\alpha/2}$, with $z_\alpha$ being the $\alpha-th$ quantile of a standard normal variable. 

Although inference based on $r_P$ is feasible, the accuracy of the results is based on asymptotic considerations, i.e., it is guaranteed when the sample size is large enough. A substantial literature warns against the risk of unreliable inferential conclusions based on the profile log-likelihood ratio statistics when the sample size is small and the asymptotic arguments do not hold (e.g., \cite{brazzale}). For example, empirical coverages of confidence intervals are lower than the nominal level and hypothesis tests can result in erroneous conclusions. In meta-analysis, recent works investigate the inaccuracy of first-order likelihood solutions when the sample size is small (\cite{guolo2012}, \cite{guolo2017}) and when the sample size within each study included in the meta-analysis is small as well (\cite{bellio}). 

When the reduced sample size cannot guarantee accuracy of asymptotic normality, the routine use of $r_P$ is discouraged and alternative solutions have been developed. The modifications of $r_P$ proposed in the literature are aimed at reducing the order of the error in approximating the standard normal (\cite{severini}, \cite{reid}). In this paper, we consider the refinement of $r_P$ given by Skovgaard's statistic \cite{skovgaard}, which improves the error of $r_P$ in approximating the standard normal distribution up to second-order. The choice is motivated by the fact that Skovgaard's statistic is well-defined for a wide class of regular problems and is computationally feasible. Moreover, the invariance with respect to interest-respecting reparameterizations is maintained. 

Skovgaard's statistic is defined as a modification of $r_P$ 
\begin{equation}\label{eqn:rskov}
\overline r_P(\psi)=  r_P(\psi) + \frac{1}{r_P(\psi)} \log{\frac{u(\psi)}{r_P(\psi)}},
\end{equation}
where $u(\psi)$ represents the correction term 
$$
u(\psi)=[S^{-1}q]_{\psi}  |\hat j |^{1/2} |\hat i |^{-1} |S| |\tilde j_{\lambda\lambda}|^{-1/2}.
$$
In the above expression, symbol $|\cdot|$ denotes the determinant, $\hat i$ and $\hat j$ are the expected information matrix and the observed information matrix, respectively, both evaluated at the maximum likelihood estimate $\hat \theta$ and $\tilde j_{\lambda\lambda}$ represent the sub-block of $j$ corresponding to the parameter vector $\lambda$ evaluated at the constrained maximum likelihood estimate $\tilde \theta$. Matrix $S$ and vector $q$ are covariances of likelihood terms. Let $\partial \ell(\theta)/\partial \theta$ denote the derivation of the log-likelihood function, i.e., the score function, with respect to $\theta$. Then
$$
S= \rm{cov}_{\theta_1} \left. \left\{\frac{\partial \ell(\theta_1)}{\partial \theta_1},  \frac{\partial \ell(\theta_2)}{\partial \theta_2}\right\} \right |_{\theta_1=\hat\theta, \theta_2=\tilde \theta}
$$
and
$$
q=\rm{cov}_{\theta_1} \left.\left\{\frac{\partial \ell(\theta_1)}{\partial \theta_1},  \ell(\theta_1)-\ell(\theta_2)\right\}\right |_{\theta_1=\hat\theta, \theta_2=\tilde \theta}.
$$
The evaluation of $S$ and $q$ at $\theta_1=\hat\theta$ and $\theta_2=\tilde\theta$ is computed after the computation of the covariance.\\
Finally, $[S^{-1}q]_{\beta_1}$ in (\ref{eqn:rskov}) is the component of the vector $S^{-1}q$ corresponding to $\psi$.

\paragraph{Example.} In order to clarify the evaluation of Skovgaard's statistic components, in the next lines we derive the expression of $S$ and $q$ in a simple framework represented by the random-effects meta-analysis model \cite{dersimonian} with equal within-study variances, where the computation is straightforward. For the more general case of meta-analysis and meta-regression, results in \cite{guolo2012} highlight the advantages of using $\overline r_P$ in place of $r_P$ in terms of accuracy of inferential conclusions for small to moderate sample sizes.

Let $Y_i$ be the measure of the effect $\upsilon$ in the $i$-th study included in a meta-analysis. Consider the linear mixed-effects model $Y_i=\upsilon_i+\epsilon_i$, where $\upsilon_i$ is the realization of a random-effect $\Upsilon_i\sim N(\upsilon, \tau^2)$, independent of $\epsilon_i \sim N(0, \sigma^2_i)$. Here, $\tau^2$ denotes the between-study variance and $\sigma^2_i$ denotes the within-study variance. Following a common and computationally convenient assumption, we assumed $\sigma^2_i$ as known. For simplicity, we focus on $\sigma_i^2=\sigma^2$. Given the above assumptions, $Y_i\sim N(\upsilon, \omega)$, where $\omega=\sigma^2+\tau^2$ for convenience. The inferential interest is on $\upsilon$, which plays the role of $\psi$ in the general setting described in Section~\ref{sec:framework}. The between-study variance $\tau^2$ represents the nuisance parameter $\lambda$.

Let $\theta=(\upsilon,\omega)^\top$ be the whole parameter vector, $\hat\theta=(\hat \upsilon, \hat \omega)^\top$ be the maximum likelihood estimate of $\theta$ and $\tilde \theta=(\upsilon, \hat \omega_\upsilon)^\top$ be the constrained maximum likelihood estimate of $\theta$ for fixed $\upsilon$. The components of the score vector are 
$$\ell_\upsilon(\theta)=\sum^n_{i=1}(y_i-\upsilon)\omega^{-1} \ \hspace{1cm} \ \ell_\omega(\theta)=0.5\sum^n_{i=1}(y_i-\upsilon)^2\omega^{-2}-0.5 \ n \ \omega^{-1} \ . $$
Ingredients of Skovgaard's statistic $\overline r(\upsilon)$ are the $2 \times 2$ matrix $S$ and the 2-dimensional vector $q$, namely,
$$
S=\left[\begin{array}{cc}
S_{\upsilon,\upsilon} & S_{\upsilon,\omega}\\
S_{\omega,\upsilon} & S_{\omega,\omega}
\end{array}\right]
\ \hspace{1cm} \ 
q=\left[\begin{array}{c}
q_\upsilon\\
q_\omega
\end{array}\right]
$$
Obtaining the components of $S$ and $q$ only requires the first three moments of a normal variable. In this way,
$$
S_{\upsilon,\upsilon}=\sum^n_{i=1}{\rm{cov}}_{\hat\theta} \left(\frac{Y_i-\hat \upsilon}{\hat \omega}, \frac{Y_i- \upsilon}{\hat\omega_\upsilon}\right)=\frac{n}{\hat\omega},
$$
$$
S_{\upsilon,\omega}=\sum^n_{i=1}{\rm{cov}}_{\hat\theta}\left(\frac{Y_i-\hat \upsilon}{\hat \omega}, \frac{(Y_i- \upsilon)^2}{2\hat\omega^2_\upsilon}\right)=\frac{n(\hat \upsilon-\upsilon)}{\hat\omega_\upsilon^2},
$$
$$
S_{\omega,\upsilon}=\sum^n_{i=1}{\rm{cov}}_{\hat\theta}\left(\frac{Y_i^2-2Y_i\hat \upsilon}{2\hat \omega^2}, \frac{Y_i}{\hat\omega_\upsilon}\right)=0,
$$
$$
S_{\omega,\omega}=\sum^n_{i=1}{\rm{cov}}_{\hat\theta}\left(\frac{(Y_i-\hat \upsilon)^2}{2\hat \omega^2}, \frac{(Y_i- \upsilon)^2}{2\hat\omega^2_\upsilon}\right)=\frac{n}{2\hat\omega_\upsilon^2},
$$
$$
q_\upsilon=-0.5\sum^n_{i=1}{\rm{cov}}_{\hat\theta}\left(\frac{Y_i-\hat \upsilon}{\hat \omega}, \frac{(Y_i- \hat \upsilon)^2}{\hat\omega} - \frac{(Y_i- \upsilon)^2}{\hat\omega_\upsilon}\right)=\frac{n(\hat \upsilon -\upsilon)}{\hat\omega_\upsilon},
$$
$$
q_\omega=-0.5\sum^n_{i=1}{\rm{cov}}_{\hat\theta}\left(\frac{(Y_i-\hat \upsilon)^2}{\hat \omega^3}, \frac{(Y_i- \hat \upsilon)^2}{\hat\omega} - \frac{(Y_i- \upsilon)^2}{\hat\omega_\upsilon}\right)=-\frac{n}{2}\left(\frac{1}{\hat \omega}-\frac{1}{\hat \omega_\upsilon}\right).
$$
The particular structure of the response examined in this example with homogeneous within-study variances represents an instance of exponential family. In this case, Skovgaard's statistic $\overline r_P$ is shown to reach a higher level of accuracy in approximating the standard normal distribution, up to third-order error in place of the second-order error (e.g., \cite{severini}).

\subsection{Skovgaard's statistic in control-rate regression}\label{sec:skovgaard}
Guolo \cite{guolo2012} investigated the use of Skovgaard's statistic in meta-analysis and meta-regression, under the classical random-effects formulation \cite{dersimonian}. This paper takes advantage of the starting results in Guolo \cite{guolo2012} to extend the usage of Skovgaard's statistic to the multivariate meta-analysis represented by control rate regression. Measurement errors on $\hat \eta_i$ and $\hat \xi_i$ are taken into account but they do not substantially affect the feasibility of the approach. 

Consider that the parameter of interest $\psi$ in Section~\ref{sec:framework} is represented by the slope $\beta_1$ of the control rate regression, so that $r_P(\psi)=r_P(\beta_1)$ and $\overline r_P(\psi)=\overline r_P(\beta_1)$. The nuisance component vector is $\lambda=(\beta_0, \mu, \tau^2, \sigma^2)^\top$. Accordingly, $S$ is a $5\times 5$ matrix with components
$$S=
\left[\begin{array}{ccccc}
S_{\beta_0,\beta_0} & S_{\beta_0,\beta_1} & S_{\beta_0,\mu}& S_{\beta_0,\tau^2} & S_{\beta_0,\sigma^2}\\
S_{\beta_1,\beta_0} & S_{\beta_1,\beta_1} & S_{\beta_1,\mu}& S_{\beta_1,\tau^2} & S_{\beta_1,\sigma^2}\\
S_{\mu,\beta_0} & S_{\mu,\beta_1} & S_{\mu,\mu}& S_{\mu,\tau^2} & S_{\mu,\sigma^2}\\
S_{\tau^2,\beta_0} & S_{\tau^2,\beta_1} & S_{\tau^2,\mu}& S_{\tau^2,\tau^2} & S_{\tau^2,\sigma^2}\\
S_{\sigma^2,\beta_0} & S_{\sigma^2,\beta_1} & S_{\sigma^2,\mu}& S_{\sigma^2,\tau^2} & S_{\sigma^2,\sigma^2}\\
\end{array}\right]
$$
and $q$ is a vector of 5 components
$$q=
\left[\begin{array}{c}
q_{\beta_0}\\
q_{\beta_1}\\
q_\mu\\
q_{\tau^2}\\
q_{\sigma^2}
\end{array}\right]
$$

The expression of the components in $S$ and $q$ is reported in the Appendix. The covariances of the likelihood terms $S$ and $q$ that give rise to the improvement of $r_P(\beta_1)$ include the measurement error correction, as the error components are taken into account both in the mean $f_i$ and in the variance/covariance matrix $V_i$, see expression (\ref{eqn:model}). Unfortunately, such a structure does not allow to write Skovgaard's components by separating higher-order terms and measurement error correction terms. Details about how to compute the components of $\overline r_P(\beta_1)$ are provided in the Supporting Web Material, Appendix A.

\section{Simulation studies}\label{sec:simulation}

Several simulation studies have been conducted to investigate the performance of Skovgaard's statistic $\overline r_P$ with respect to the signed profile log-likelihood ratio statistic $r_P$ in terms of accuracy of inferential results about $\beta_1$. Both the approaches are compared to the usual weighted least squares regression. 

Data have been simulated with a two-step procedure. In the first step the number of events within each study included in the meta-analysis are generated. In the second step, the generated data are used to produce the outcome measure of interest in the treated group and in the control group. 
We consider $\eta_i$ and $\xi_i$ as the log event rate in the treatment group and in the control group, respectively. Their observed versions are $\hat \eta_i=\log (y_i/n_i)$ and $\hat \xi_i=\log (x_i/m_i)$, respectively, where $y_i$ and $n_i$ are the number of events and the total number of person-years 
in study $i$ in the treatment group and $x_i$ and $m_i$ are the number of events and the total number of person-years 
in study $i$ in the control group, respectively. The variance of a log event rate is given by the inverse of the number of observed events, so that 
\begin{equation}\label{eqn:gamma}
\Gamma_i=
\left[
\begin{array}{cc}
y_i^{-1} & 0 \\
0 & x_i^{-1}
\end{array}
\right],
\end{equation}
where the null covariance is a consequence of the event rates calculated on independent groups (e.g., \cite{arends}).
For fixed number of studies $n$, the number of events in each study included in the meta-analysis $y_i$ and $x_i$ are simulated from the distributions $Y_i\sim {\rm Poisson}(n_i e^{\eta_i})$ and $X_i \sim {\rm Poisson}(m_i e^{\xi_i})$ \cite{arends}. Quantities $n_i$ and $m_i$ in each study $i$ are generated from a Uniform variable on $[100, 5000]$. Values of $\xi_i$ are simulated from a $N(\mu,\sigma^2)$ with $\mu$ and $\sigma^2$ specified as described in the next lines and values of $\eta_i$ are obtained from the regression line (\ref{eqn:model1}).
The number of studies $n$ is small to moderate, with values $n \in \{5, 10, 20\}$. The square root $\tau$ of the variance component $\tau^2$ assumes increasing values in a grid from 0.3 to 2, while the variance component $\sigma^2$ is initially set equal to 1. The performance of the methods for varying $\sigma^2$ will be examined later.  Parameters $\beta_0, \beta_1,\mu$ are chosen in order to reflect scenarios with reducing event rate, namely, scenario 1 with $(\beta_0, \beta_1, \mu)^\top=(0, 1, 1)^\top$, scenario 2 with $(\beta_0, \beta_1, \mu)^\top=(-1.5, 1, -0.5)^\top$, scenario 3 with $(\beta_0, \beta_1, \mu)^\top=(-1.5, 1, -2.5)^\top$, scenario 4 with $(\beta_0, \beta_1, \mu)^\top=(-3, 1, -2)^\top$.

The simulation experiment has been repeated 1,000 times for each scenario and for each combination of $\tau$ and $n$. The methods are compared in terms of empirical coverage probabilities of confidence intervals for $\beta_1$ at nominal level 0.95. When using the weighted least squares regression, the Wald-type confidence interval is considered. Likelihood maximisation, based on the Nelder and Mead algorithm \cite{nelder}, employs the weighted least squares estimates as starting values.  

Simulation results are reported in Figure~1 for scenario 1. Skovgaard's statistic provides empirical coverages of confidence intervals very close to the nominal level, independently of the sample size $n$ and the amount of variance $\tau^2$. The improvement provided by the method over alternative approaches is pronounced and more evident in case of small $n$ as well as large $\tau^2$. See, for example, the results for $n=5$ and for $\tau=2.0$. 
Relying on first-order likelihood inference turns out in confidence intervals with empirical coverage probabilities substantially lower than the nominal level when the sample size is small. Differences with respect to Skovgaard's statistic reduce as the sample size increases, as expected from a theoretical point of view.
Unsurprisingly, the weighted least squares regression shows a pronounced unsatisfactory behaviour, as a consequence of not accounting for measurement errors. The empirical coverage probabilities notably underestimate the nominal 95\% level, more seriously as the amount of between-study heterogeneity increases. Results for scenarios 2, 3 and 4 are reported in the Supporting Web Material, Appendix B. They substantially confirm the previous findings. Skovgaard's statistic globally maintains satisfactory empirical coverages of confidence intervals over alternatives. A small deviation from the target level emerging for small $\tau$ when the event rate is very low disappears as the sample size increases.

 \begin{figure}[h]
\begin{center}
\includegraphics[width=5.5in]{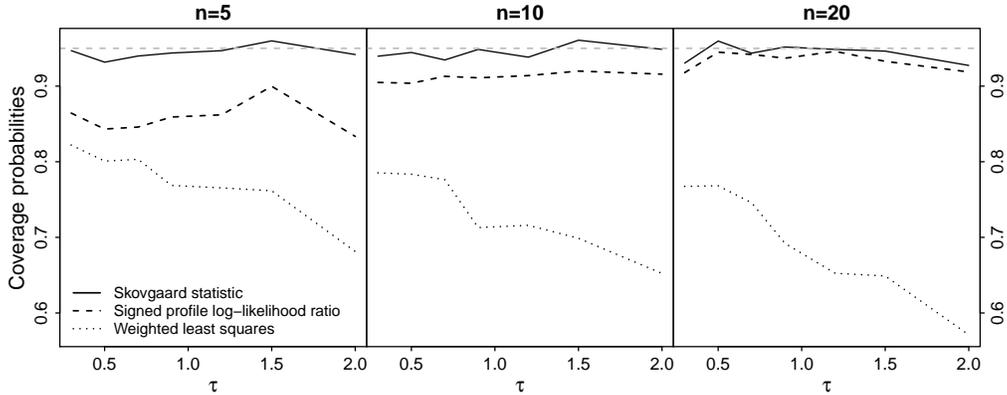}
\caption{Empirical coverage probabilities of the nominally 95\% confidence interval for $\beta_1$, when $(\beta_0, \beta_1, \mu)^\top=(0, 1, 1)^\top$, under increasing sample size $n$ and square root $\tau$ of the variance component $\tau^2$. Variance component $\sigma^2=1$. The plotted curves correspond to Skovgaard's statistic (solid), the signed profile log-likelihood ratio statistic (dashed), the weighted least squares approach (dotted). The dashed, grey horizontal line is the nominal level.}
\end{center}
\end{figure}

Additional simulation studies have been performed to evaluate the impact of varying $\sigma^2$ and the corresponding results are reported in the Supporting Web Material, Appendix B. The studies examine the performance of the methods when $\tau$ is fixed at $1.2$ and $\sigma$ assumes increasing values in a grid from 0.3 to 1.5.  
Results for the four scenarios with reducing event rate again highlights advantages of relying on Skovgaard's statistic. Empirical coverage probabilities are closer to the target level than alternatives, with emphasis in case of small $n$. See the substantial discrepancies between Skovgaard's statistic and the first-order counterpart $r_P$ when $n=5$. The unsatisfactory performance of the weighted least squares regression persists whichever the scenario. Globally, the underestimation of the nominal level becomes worse as the value of $\sigma$ increases.

\section{Example}\label{sec:example}
Hoes et al. \cite{hoes} consider a meta-analysis of 12 studies about the efficacy of a drug treatment compared to placebo or no treatment to prevent death for cardiovascular reasons in middle-aged patients with mild to moderate hypertension. The available information is in terms of the number of events and the total number of person-years per group, as reported in Table~\ref{tab:hoes}.

\begin{table}[ht]
\centering
\begin{tabular}{rrrrr}
  \hline
 Study  &\multicolumn{2}{c}{Treatment group}&\multicolumn{2}{c}{Control group}\\
 & Deaths & Person-years & Deaths & Person-years \\ 
  \hline
1 & 10 & 595.2 & 21 & 640.2 \\ 
  2 & 2 & 762.0 & 0 & 756.5 \\ 
  3 & 54 & 5635.0 & 70 & 5600.0 \\ 
  4 & 47 & 5135.0 & 63 & 4960.0 \\ 
  5 & 53 & 3760.0 & 62 & 4210.0 \\ 
  6 & 10 & 2233.0 & 9 & 2084.5 \\ 
  7 & 25 & 7056.1 & 35 & 6824.0 \\ 
  8 & 47 & 8099.0 & 31 & 8267.0 \\ 
  9 & 43 & 5810.0 & 39 & 5922.0 \\ 
  10 & 25 & 5397.0 & 45 & 5173.0 \\ 
  11 & 157 & 22162.7 & 182 & 22172.5 \\ 
  12 & 92 & 20885.0 & 72 & 20645.0 \\
   \hline
\end{tabular}\caption{Number of deaths and total number of person-years in the treatment and control group of mild to moderate hypertension middle-aged patients in the meta-analysis of Hoes et al. \cite{hoes}.}\label{tab:hoes}
\end{table}

Let $\eta_i$ and $\xi_i$ denote the log mortality rate for the $i$-th treatment group and control group, respectively. The slope of the regression line (\ref{eqn:model1}) is thus tested against one, i.e., the slope on no-effect line, see Arends et al. \cite{arends}. To this aim, consider the observed error-prone $\hat \eta_i$ and $\hat \xi_i$ evaluated as the logarithm of the number of deaths over the total number of person-years in the treatment and in the control group, respectively. The associated variance/covariance matrix $\Gamma_i$ follows expression (\ref{eqn:gamma}). 
The maximum likelihood estimate of $\beta_1$ is equal to 0.69, with standard error 0.08. Testing for $\beta_1$ equal to one using the first-order likelihood approach results in an indication of effect of the drug treatment in reducing the risk of mortality, as the signed profile log-likelihood ratio statistic is $r_P(\beta_1)=-2.34$, with an associated p-value equal to 0.02. The associated 95\% confidence interval for $\beta_1$ obtained using the standard normal approximation for $r_P$ is equal to $(0.45; 0.93)$. The result is in line with the analysis provided by Arends et al. \cite{arends} through a Bayesian approach. When considering Skovgaard's statistic, instead, results change. In this case, in fact, $\overline r_P(\beta_1)=-1.27$, with an associated p-value equal to 0.20. The associated 95\% confidence interval for $\beta_1$ using the standard normal approximation of $\overline r_P$ is $(0.38; 1.13)$. 
Thus, taking into account the small number of studies included in the analysis through a higher-order likelihood solution change first-order results and conclude for no effect of the drug treatment to prevent death for cardiovascular reasons.

\section{Concluding remarks}\label{sec:conclusion}
This paper considered likelihood inference in control rate regression accounting for the presence of measurement error affecting the outcome risk measure in the treatment group and in the control group. Attention has been paid to situations with a small number of studies, where first-order results based on the log-likelihood ratio statistic can be substantially inaccurate. In order to avoid misleading inferential conclusions, we suggested to base inference on Skovgaard's statistic, which improves to the second-order the accuracy in approximating the standard normal distribution. The simulation experiments show that the empirical coverage probabilities of confidence intervals for $\beta_1$ based on Skovgaard's statistic tend to be closer to the nominal level than those derived from the log-likelihood ratio statistic. The improvements are more evident when the number of studies included in the meta-analysis is small, e.g., $n=5$, and with increasing variance $\tau^2$. The gain in accuracy is reached with no appreciable computational effort, as the evaluation of Skovgaard's statistic components has a complexity comparable to that of computing the expected information matrix.

The simulation study and the data analysis have been implemented using the \texttt R programming language \cite{r}. The \texttt R code for computing Skovgaard's statistic is provided as Supporting Web Material. Moreover, Appendix C in the Supporting Web Material includes an illustration about how to use the software in order to implement Skovgaard's statistic in control rate regression. 

Likelihood inference performed in this paper, using either first-order or higher-order solutions, considers the approximate normal distribution (\ref{eqn:mem}) for the measurement error, which assures the likelihood function being in closed-form. This means that, when necessary, the correction that adds 0.5 to the number of events equal to zero is applied, as, for example, to avoid inadmissible values of the estimated log event rate and its variance. The correction is not needed when using the exact measurement error structure at the price of obtaining the likelihood function not in closed-form, see \cite{arends}. In this case, Skovgaard's statistic is still evaluable, but the order of the approximation of $\overline r_P$ to the standard normal is not known, as a consequence of the numerical integration. Nevertheless, experimental studies in Guolo et al. \cite{guolo2006} shows that a good performance of Skovgaard's statistic with respect to the first-order solution is maintained in random-effects models, when the sample size is large. In addition, empirical investigations performed with reference to the data analysis in Section~\ref{sec:example} with different correction values show that the 0.5 correction does not impact the results. 

In this paper we follow the classical assumption considering the within-study variances as known and equal to the estimate provided by the studies included in the meta-analysis. Such an approach is justified in the common case of large within-study sample size. In cases where the assumption does not hold, a proper analysis should account for the uncertainty in measuring the within-study variances. In the classical random-effects meta-analysis framework, Bellio and Guolo \cite{bellio} investigated a likelihood approach which includes an extra component accounting for the additional source of variability. Although a similar extension would be possible in control rate regression, the resulting likelihood function is expected not to be in closed-form. As previously mentioned, in such a case the order of the approximation of Skovgaard's statistic to the standard normal is not known, as a consequence of numerical integration. 

Although we considered the approximate model (\ref{eqn:mem}), the performance of the signed profile log-likelihood ratio statistic $r_P$ based on the exact likelihood function has been investigated through simulation. Numerical integration used a Gauss-Hermite quadrature with 50 to 100 nodes and pruning at level 20\%. In the scenarios examined in this paper, the application of the method was challenging given computational drawbacks which made the approach unappealing. Substantial computational difficulties emerged manly in terms of non-convergence of the optimisation algorithm, with non-positive definite variance/covariance matrix or unreliable parameter estimates on the boundary of the parameter space. This gave rise to large failure rates, up to 50\% for extreme cases with $n=5$ and large between-study variance. Modifications to the integral evaluation, such as adaptive quadratures, or modifications to the optimization algorithm, such as changes in the optimizer or in the starting values, did not succeed in reducing the convergence problems. 
On the other hand, when the method converges, then results in terms of empirical coverage of confidence interval at nominal level 95\% for $r_P$ are comparable to those obtained under the approximate normal model (\ref{eqn:mem}). Again, the approximation of $r_P$ to the standard normal distribution is poor and second-order Skovgaard's statistic remains a preferable solution.

\section*{Supporting information}
The Supporting Web Material includes the derivation of the Skovgaard's statistic components (Appendix A), additional simulation results (Appendix B), the analysis of the data in Hoes et al. \cite{hoes} (Appendix C), the \texttt R code for applying Skovgaard's statistic.

\section*{Acknowledgments}
This work was supported by a grant from the University of Padova (Progetti di Ricerca di Ateneo 2015, CPDA153257). The Author is grateful to Prof. Ruggero Bellio and to Prof. Cristiano Varin for helpful discussions.

\appendix

\section{Components of Skovgaard's statistic}
Consider the notation in Section~\ref{sec:skovgaard}. Denote by $f_i$ the mean vector of $(\hat \eta_i, \hat \xi_i)^\top$ and by $V_i$ the associated variance/covariance matrix in (\ref{eqn:model}). A subfix indicates the derivation with respect to each component of $\theta$. A "hat" and a "tilde" indicate the evaluation of a vector or a matrix with respect to $\hat \theta$ and $\tilde \theta$, respectively. The components of $S$ are
$$
S_{\beta_j,\beta_k}= \sum^n_{i=1} \left\{ \frac{1}{2} {\rm trace}\left(\hat V^{-1}_{\beta_j}\hat V_i \tilde V^{-1}_{\beta_k}\hat V_i\right) + \hat f_{i,\beta_j}^\top \tilde V^{-1}_{\beta_k} \left(\tilde f_i-\hat f_i\right) + \hat f_{\beta_k}\tilde V^{-1}_i \tilde f_{\beta_k} \right \}, \ j,k=0,1,
$$
$$
S_{\beta_j,\mu}= \sum^n_{i=1} \hat f^\top_{i,\beta_j} \tilde V^{-1}_i \tilde f_{i,\mu}, \ j=0,1,
$$
$$
S_{\mu,\mu} = \sum^n_{i=1} \hat f^\top_{i,\mu} \tilde V^{-1}_i \tilde f_{i,\mu},
$$
$$
S_{\beta_j, \psi_k} =\sum^n_{i=1} \left\{ \frac{1}{2} {\rm trace}\left(\hat V^{-1}_{i,\beta_j}\hat V_i \tilde V^{-1}_{i,\psi_k}\hat V_i\right) + \hat f_{i,\beta_j}^\top \tilde V^{-1}_{i,\psi_k} \left(\tilde f_i-\hat f_i\right)\right \},  \ j=0,1, \ \psi_k \in \{\tau^2,\sigma^2\},
$$
$$
S_{\mu, \psi_k} =\sum^n_{i=1} \left\{ \frac{1}{2} {\rm trace}\left(\hat V^{-1}_{i,\mu}\hat V_i \tilde V^{-1}_{i,\psi_k}\hat V_i\right) + \hat f_{i,\mu}^\top \tilde V^{-1}_{i,\psi_k} \left(\tilde f_i-\hat f_i\right)\right \},  \ \psi_k \in \{\tau^2,\sigma^2\}, \ \psi_k \in \{\tau^2,\sigma^2\},
$$
$$
S_{\psi_j, \psi_k} =\frac{1}{2} \sum^n_{i=1} {\rm trace}\left(\hat V^{-1}_{i,\psi_j}\hat V_i \tilde V^{-1}_{i,\psi_k}\hat V_i\right), \ \psi_j,\psi_k \in \{\tau^2,\sigma^2\},
$$
$$
S_{\mu, \beta_j} = \sum^n_{i=1}\left( \hat f^\top_{i,\mu} \tilde V^{-1}_i \tilde f_{i,\beta_j} + \hat f^\top_{i,\mu} \tilde V^{-1}_{i,\beta_j} \tilde f_i - \hat f_i^\top \tilde V^{-1}_{i,\beta_j} \hat f_{i,\mu}\right), \ j=0,1
$$
$$
S_{\psi_j, \beta_k} = \frac{1}{2} \sum^n_{i=1} {\rm trace}\left(\hat V^{-1}_{i,\psi_j} \hat V_i \tilde V^{-1}_{i,\beta_k}\hat V_i\right), \ \psi_j\in \{\tau^2,\sigma^2\}, k=0,1
$$
$$
S_{\psi_j, \mu} =0, \ \psi_j\in \{\tau^2,\sigma^2\}.
$$
Similarly, the components of $q$ are
$$
q_{\beta_j}= \sum^n_{i=1} \left[ \frac{1}{2} {\rm trace}\left\{\hat V^{-1}_{i,\beta_j}\hat V_i \left(\hat V^{-1}_i - \tilde V^{-1}_i\right) \hat V_i\right\} + \hat f_{i,\beta_j}^\top \tilde V^{-1}_i \left(\hat f_i-\tilde f_i\right) \right], \ j=0,1,
$$
$$
q_\mu =  \sum^n_{i=1} \left[ \frac{1}{2} {\rm trace}\left\{\hat V^{-1}_{i,\mu}\hat V_i \left(\hat V^{-1}_i - \tilde V^{-1}_i\right) \hat V_i\right\} + \hat f_{i,\mu}^\top \tilde V^{-1}_i (\hat f_i-\tilde f_i) \right]
$$
and
$$
q_{\psi_j} = \frac{1}{2} \sum^n_{i=1} \left\{{\rm trace} \left(\hat V^{-1}_{\psi_j} \hat V_i\right) - {\rm trace}\left(\hat V^{-1}_{\psi_j} \hat V_i \tilde V^{-1}_i \hat V_i\right)  \right\}, \ \psi_j \in \{\tau^2,\sigma^2\}.
$$
Details about how to compute the components of $\overline r_P(\beta_1)$ are provided in the Supporting Web Material, Appendix A.

\newpage
\begin{center}\LARGE{
Web-based Supporting Materials for "Improving likelihood-based inference in control rate regression"
\\
by \\
Annamaria Guolo}
\end{center}

\vspace{2cm}

\appendix
\section*{Web Appendix A: Derivation of Skovgaard's statistic}
Given the framework described in Section~2 of the paper, the log-likelihood function $\ell(\theta)$ for the whole parameter vector $\theta$ is 
$$
\ell(\theta)\propto -\frac{1}{2}\sum^n_{i=1} \log |V_i| - \frac{1}{2}\sum^n_{i=1} (y_i-f_i)^\top V_i^{-1} (y_i-f_i),
$$
where $y_i=(\hat \eta_i, \hat \xi_i)^\top$ is the observed value of the random vector $Y_i$ with mean vector $f_i$ and variance/covariance matrix $V_i$, following the notation in Section~3.1 of the paper. 
The score vector
$$
\ell_\theta(\theta)=
\left[
\begin{array}{l}
\ell_{\beta_0} (\theta)\\
\ell_{\beta_1} (\theta)\\
\ell_{\mu} (\theta)\\
\ell_{\tau^2} (\theta)\\
\ell_{\sigma^2} (\theta)\\
\end{array}\right]
$$
has components
\begin{eqnarray}\nonumber
\ell_{\beta_j}(\theta)&=&-\frac{1}{2} \sum^n_{i=1} {\rm trace}\left(V_i^{-1} V_{i,\beta_j}\right)\\ \nonumber
&& -\frac{1}{2} \sum^n_{i=1} \left(
y_i^\top V^{-1}_{i,\beta_j}y_i - 2 f^\top_{i,\beta_j} V^{-1}_i y_i - 2 f_i^\top V^{-1}_{i,\beta_j} y_i +2f_{i,\beta_j} V^{-1}_i f_i + f^\top_i V^{-1}_{i,\beta_j}f_i\right), j=0,1,
\end{eqnarray}
$$
\ell_\mu(\theta)= \sum^n_{i=1} f^\top_{i,\mu}V_i^{-1}(y_i-f_i)
$$
and
$$
\ell_{\psi_j}(\theta) = -\frac{1}{2}\sum^n_{i=1} {\rm trace} \left(V_i^{-1}V_{i,\psi_j} \right)-\frac{1}{2} \sum^n_{i=1} \left( 
y_i^\top V^{-1}_{i,\psi_j}y_i - 2f_i^\top V^{-1}_{i,\psi_j}y_i + f_i^\top V^{-1}_{i,\psi_j}f_i\right), \ \psi_j \in \{\tau^2,\sigma^2\}.
$$
The expected information matrix $$
i(\theta)=
\left[
\begin{array}{lllll}
i_{\beta_0\beta_0}(\theta)  & i_{\beta_0\beta_1}(\theta)& i_{\beta_0\mu}(\theta) & i_{\beta_0\tau^2}(\theta) & i_{\beta_0\sigma^2}(\theta) \\
i_{\beta_0\beta_1}(\theta)& i_{\beta_1\beta_1}(\theta) &  i_{\beta_1\mu}(\theta) &i_{\beta_1\tau^2}(\theta) &i_{\beta_1\sigma^2}(\theta) \\
i_{\beta_0\mu}(\theta) &i_{\beta_1\mu}(\theta)&i_{\mu\mu}(\theta) &i_{\mu\tau^2}(\theta) & i_{\mu\sigma^2}(\theta)\\
i_{\beta_0\tau^2}(\theta)& i_{\beta_1\tau^2}(\theta)& i_{\mu\tau^2}(\theta)& i_{\tau^2\tau^2}(\theta) & i_{\tau^2\sigma^2}(\theta)\\
i_{\beta_0\sigma^2}(\theta) &i_{\beta_1\sigma^2}(\theta)&i_{\mu\sigma^2}(\theta)&i_{\tau^2\sigma^2}(\theta)&i_{\sigma^2\sigma^2}(\theta)\\
\end{array}\right]
$$
has generic component
$$
i_{\theta_j\theta_k}= \frac{1}{2} \sum^n_{i=1} {\rm trace} \left(
V^{-1}_{i,\theta_k}V_{i,\theta_j} + V^{-1}_i V_{i, \theta_j\theta_k} - V^{-1}_{i,\theta_j}V_{i,\theta_j\theta_k}V^{-1}_{i,\theta_j}V_i \right) + \sum^n_{i=1} f_{i,\theta_j} V^{-1}_i f_{i,\theta_k}, \ \theta_j,\theta_k \in \theta,
$$
where $V_{i, \theta_j\theta_k}$ denotes the second derivative of $V_i$ with respect to $\theta_j,\theta_k\in \theta$.
In order to derive the components of $S$ and $q$, consider that
$$
{\rm cov}\left(Y_i^\top \hat V^{-1}_{i,\theta_j}Y_i, Y_i^\top \tilde V_{i,\theta_k}Y_i\right) = {\rm trace}\left(\hat V^{-1}_{i,\theta_j} \hat V_i \tilde V^{-1}_{i,\theta_k}\hat V_i\right) + 4 \hat f_i^\top \hat V^{-1}_{i,\theta_j}\hat V_i \tilde V^{-1}_{i,\theta_k} \hat f_i,
$$
$$
{\rm cov}\left(Y_i^\top \hat V^{-1}_{i,\theta_j}Y_i, Y_i^\top \tilde V_{i,\theta_k}Y_i\right) = 2f_i^\top \hat V^{-1}_{i,\theta_j} \hat V_i \tilde V_i^{-1} \tilde f_{i,\theta_j}
$$
and
$$
{\rm cov}\left(\hat f_{i,\theta_j} \hat V^{-1}_i Y_i, \tilde f_i^\top \tilde V^{-1}_{i,\theta_k} Y_i\right) = \hat f^\top_{i,\theta_j} \hat V^{-1}_i  \hat V_i  \tilde V^{-1}_{i,\theta_k} \tilde f_i,
$$
for $\theta_j,\theta_k \in \theta$.

Then,
\begin{eqnarray}\nonumber
S_{\beta_j,\beta_k}(\theta)&=&\left. {\rm cov}\left\{\ell_{\beta_j}(\theta_1), \ell_{\beta_k}(\theta_2)\right\} \right |_{\theta_1=\hat\theta, \theta_2=\tilde \theta}\\ \nonumber
&=& \frac{1}{4} {\rm cov} \sum_{i=1}^n \Big(
Y_i \hat V^{-1}_{i,\beta_j} Y_i - 2 \hat f^\top_{i,\beta_j} \hat V_i^{-1}Y_i - 2 \hat f_i \hat V^{-1}_{i,\beta_j} Y_i, Y_i^\top \tilde V_{i,\beta_k}^{-1} Y_i -2\tilde f^\top_{i,\beta_k} \tilde V^{-1}_i Y_i \\ \nonumber
&& -2 \tilde f^\top_i \tilde V^{-1}_{i,\beta_k}Y_i\Big)\\ \nonumber
&=&
\sum^n_{i=1} \left\{ \frac{1}{2} {\rm trace}\left(\hat V^{-1}_{i,\beta_j}\hat V_i \tilde V^{-1}_{i,\beta_k}\hat V_i\right) + \hat f_{i,\beta_j}^\top \tilde V^{-1}_{i,\beta_k} \left(\tilde f_i-\hat f_i\right) + \hat f_{i,\beta_k}\tilde V^{-1}_i \tilde f_{i,\beta_k} \right \}, \ j,k=0,1
\end{eqnarray}
\begin{eqnarray}\nonumber
S_{\beta_j,\mu}(\theta)&=&\left. {\rm cov}\left\{\ell_{\beta_j}(\theta_1), \ell_{\mu}(\theta_2)\right\} \right |_{\theta_1=\hat\theta, \theta_2=\tilde \theta}\\ \nonumber
&=& \frac{1}{2} {\rm cov} \sum^n_{i=1} \left( Y_i^\top \hat V^{-1}_{i,\beta_j} Y_i -\hat f^\top_{i,\beta_j}\hat V_i^{-1} Y_i -2 \hat f_i \hat V^{-1}_{i,\beta_j} Y_i, \tilde f^\top_{i,\mu} \tilde V^{-1}_i Y_i\right)\\ \nonumber
&=& \sum^n_{i=1} \hat f^\top_{i,\beta_j} \tilde V^{-1}_i \tilde f_{i,\mu}, \ j=0,1
\end{eqnarray}
\begin{eqnarray}\nonumber
S_{\mu,\mu}(\theta)&=& \left. {\rm cov}\left\{\ell_{\mu}(\theta_1), \ell_{\mu}(\theta_2)\right\} \right |_{\theta_1=\hat\theta, \theta_2=\tilde \theta}\\ \nonumber
&=& {\rm cov} \sum^n_{i=1} \left(\hat f^\top_{i,\mu} \hat V^{-1}_i Y_i, \tilde f^\top_{i,\mu} \tilde V^{-1}_i Y_i \right)\\ \nonumber
&=& \sum^n_{i=1} \hat f^\top_{i,\mu} \tilde V^{-1}_i \tilde f_{i,\mu}
\end{eqnarray}
\begin{eqnarray}\nonumber
S_{\beta_j,\psi_k}(\theta)&=&\left. {\rm cov}\left\{\ell_{\beta_j}(\theta_1), \ell_{\psi_k}(\theta_2)\right\} \right |_{\theta_1=\hat\theta, \theta_2=\tilde \theta}\\ \nonumber
&=&\frac{1}{2} {\rm cov} \sum^n_{i=1}  \left(
Y_i^\top \hat V^{-1}_{i,\beta_j} Y_i -2 \hat f^\top_{i,\beta_j} \hat V^{-1}_i Y_i -2 \hat f^\top_i \hat V^{-1}_{i,\beta_j} Y_i, Y_i^\top \tilde V^{-1}_{i,\psi_k} Y_i - 2 \tilde f_i^\top \tilde V^{-1}_{i,\psi_k} Y_i \right)
 \\ \nonumber
&=&\sum^n_{i=1} \left\{ \frac{1}{2} {\rm trace}\left(\hat V^{-1}_{i,\beta_j}\hat V_i \tilde V^{-1}_{i,\psi_k}\hat V_i\right) + \hat f_{i,\beta_j}^\top \tilde V^{-1}_{i,\psi_k} \left(\tilde f_i-\hat f_i\right)\right \},  \ j=0,1, \ \psi_k \in \{\tau^2,\sigma^2\}
\end{eqnarray}
\begin{eqnarray}\nonumber
S_{\mu,\psi_k}(\theta)&=&\left. {\rm cov}\left\{\ell_{\mu}(\theta_1), \ell_{\psi_k}(\theta_2)\right\} \right |_{\theta_1=\hat\theta, \theta_2=\tilde \theta}\\ \nonumber
&=&-\frac{1}{2} {\rm cov} \sum^n_{i=1}  \left(\hat f_{i,\mu} \hat V^{-1}_iY_i, Y_i^\top \tilde V^{-1}_i Y_i -2 \tilde f_i^\top \tilde V^{-1}_{i,\psi_k} Y_i\right) \\ \nonumber
&=&\sum^n_{i=1} \left\{ \frac{1}{2} {\rm trace}\left(\hat V^{-1}_{i,\mu}\hat V_i \tilde V^{-1}_{i,\psi_k}\hat V_i\right) + \hat f_{i,\mu}^\top \tilde V^{-1}_{i,\psi_k} \left(\tilde f_i-\hat f_i\right)\right \},  \ \psi_k \in \{\tau^2,\sigma^2\}
\end{eqnarray}

\begin{eqnarray}\nonumber
S_{\psi_j,\psi_k}(\theta)&=&\left. {\rm cov}\left\{\ell_{\psi_j}(\theta_1), \ell_{\psi_k}(\theta_2)\right\} \right |_{\theta_1=\hat\theta, \theta_2=\tilde \theta}\\ \nonumber
&=&\frac{1}{4} {\rm cov} \sum^n_{i=1}  \left(Y_i^\top\hat V^{-1}_{i,\psi_j}Y_i - 2 \hat f_i^\top \hat V^{-1}_{i,\psi_j} Y_i, Y_i^\top \tilde V^{-1}_{i,\psi_k} Y_i - 2 \tilde f_i^\top \tilde V^{-1}_{i,\psi_k} Y_i\right) \\ \nonumber
&=& \frac{1}{2} \sum^n_{i=1} {\rm trace}\left(\hat V^{-1}_{i,\psi_j}\hat V_i \tilde V^{-1}_{i,\psi_k}\hat V_i\right), \ \psi_j,\psi_k \in \{\tau^2,\sigma^2\}
\end{eqnarray}

\begin{eqnarray}\nonumber
S_{\mu,\beta_j}(\theta)&=&\left. {\rm cov}\left\{\ell_{\mu}(\theta_1), \ell_{\beta_k}(\theta_2)\right\} \right |_{\theta_1=\hat\theta, \theta_2=\tilde \theta}\\ \nonumber
&=& -\frac{1}{2}{\rm cov} \sum^n_{i=1}  \left(
\hat f^\top_{i,\mu} \hat V^{-1}_i Y_i, Y_i^\top \tilde V^{-1}_{i,\beta_j}Y_i -2 \tilde f_{i,\beta_j}^\top \tilde V^{-1}_i Y_i -2 \tilde f_i \tilde V^{-1}_{i,\beta_j}Y_i\right)\\ \nonumber
&=&\sum^n_{i=1}\left( \hat f^\top_{i,\mu} \tilde V^{-1}_i \tilde f_{i,\beta_j} + \hat f^\top_{i,\mu} \tilde V^{-1}_{i,\beta_j} \tilde f_i - \hat f_i^\top \tilde V^{-1}_{i,\beta_j} \hat f_{i,\mu}\right), \ j=0,1
\end{eqnarray}

\begin{eqnarray}\nonumber
S_{\psi_j,\beta_k}(\theta)&=&\left. {\rm cov}\left\{\ell_{\psi_j}(\theta_1), \ell_{\beta_k}(\theta_2)\right\} \right |_{\theta_1=\hat\theta, \theta_2=\tilde \theta}\\ \nonumber
&=& \frac{1}{4} {\rm cov} \sum^n_{i=1}  \left(
Y_i^\top \hat V^{-1}_{i,\psi_j}Y_i -2 \hat f_i^\top \hat V^{-1}_{i,\psi_j}Y_i, Y_i^\top \tilde V^{-1}_{i,\beta_k}Y_i -2\tilde f^\top_{i,\beta_k}\tilde V^{-1}_i Y_i - 2 \tilde f_i^\top \tilde V^{-1}_{i,\beta_k}Y_i\right)\\ \nonumber
&=&\frac{1}{2} \sum^n_{i=1} {\rm trace}\left(\hat V^{-1}_{i,\psi_j} \hat V_i \tilde V^{-1}_{i,\beta_k}\hat V_i\right), \psi_j\in \{\tau^2,\sigma^2\}, k=0,1
\end{eqnarray}

\begin{eqnarray}\nonumber
S_{\psi_j,\mu}(\theta)&=&\left. {\rm cov}\left\{\ell_{\psi_j}(\theta_1), \ell_{\mu}(\theta_2)\right\} \right |_{\theta_1=\hat\theta, \theta_2=\tilde \theta}\\ \nonumber
&=& -\frac{1}{2} {\rm cov} \sum^n_{i=1}  \left(Y_i^\top \hat V^{-1}_{i,\psi_k}Y_i -2 \hat f_i^\top \hat V^{-1}_{i,\psi_k} Y_i, \tilde f^\top_{i,\mu} \tilde V_i^{-1} Y_i\right)\\ \nonumber
&=&0, \ \psi_j\in \{\tau^2,\sigma^2\}
\end{eqnarray}

\begin{eqnarray}\nonumber
q_{\beta_j}(\theta)&=& 
\rm{cov}\left.\left\{\ell_{\beta_j}(\theta_1) ,\ell(\theta_1)-\ell(\theta_2)\right\}\right |_{\theta_1=\hat\theta, \theta_2=\tilde \theta}
\\ \nonumber
&=& \frac{1}{4} {\rm cov}\sum^n_{i=1} \Big(Y_i^\top \hat V^{-1}_{i,\beta_j}Y_i -2 \hat f^\top_{i,\beta_j}\hat V^{-1}_i Y_i -2 \hat f^\top_i \hat V^{-1}_{i,\beta_j}Y_i, Y_i^\top \hat V^{-1}_i Y_i \\ \nonumber
&&- 2 f_i^\top \hat V^{-1}_i Y_i - Y_i^\top \tilde V^{-1}_iY_i + 2 \tilde f_i \tilde V_i^{-1} Y_i\Big)\\ \nonumber
&=& \sum^n_{i=1} \left[ \frac{1}{2} {\rm trace}\left\{\hat V^{-1}_{i,\beta_j}\hat V_i \left(\hat V^{-1}_i - \tilde V^{-1}_i\right) \hat V_i\right\} + \hat f_{i,\beta_j}^\top \tilde V^{-1}_i \left(\hat f_i-\tilde f_i\right) \right], \ j=0,1
\end{eqnarray}

\begin{eqnarray}\nonumber
q_{\mu}(\theta)&=& 
\rm{cov}\left.\left\{\ell_{\mu}(\theta_1) ,\ell(\theta_1)-\ell(\theta_2)\right\}\right |_{\theta_1=\hat\theta, \theta_2=\tilde \theta}
\\ \nonumber
&=& \frac{1}{4} {\rm cov} \sum^n_{i=1} \Big(
Y_i^\top \hat V^{-1}_{i,\mu}Y_i -2 \hat f^\top_{i,\mu}\hat V^{-1}_i Y_i -2 \hat f^\top_i \hat V^{-1}_{i,\mu}Y_i, Y_i^\top \hat V^{-1}_i Y_i \\ \nonumber
&& - 2 f_i^\top \hat V^{-1}_i Y_i - Y_i^\top \tilde V^{-1}_iY_i + 2 \tilde f_i \tilde V_i^{-1} Y_i\Big)\\ \nonumber
&=&\sum^n_{i=1} \left[ \frac{1}{2} {\rm trace}\left\{\hat V^{-1}_{i,\mu}\hat V_i \left(\hat V^{-1}_i - \tilde V^{-1}_i\right) \hat V_i\right\} + \hat f_{i,\mu}^\top \tilde V^{-1}_i \left(\hat f_i-\tilde f_i\right) \right]
\end{eqnarray}

\begin{eqnarray}\nonumber
q_{\psi_j}(\theta)&=& 
\rm{cov} \left.\left\{\ell_{\psi_j}(\theta_1) ,\ell(\theta_1)-\ell(\theta_2)\right\}\right |_{\theta_1=\hat\theta, \theta_2=\tilde \theta}
\\ \nonumber
&=& \frac{1}{4} {\rm cov} \sum^n_{i=1} \Big(
Y_i^\top \hat V^{-1}_{i,\psi_j}Y_i -2 \hat f^\top_{i,\psi_j}\hat V^{-1}_i Y_i -2 \hat f^\top_i \hat V^{-1}_{i,\mu}Y_i, Y_i^\top \hat V^{-1}_i Y_i \\ \nonumber
&& - 2 f_i^\top \hat V^{-1}_i Y_i - Y_i^\top \tilde V^{-1}_iY_i + 2 \tilde f_i \tilde V_i^{-1} Y_i\Big)\\ \nonumber
&=&\frac{1}{2} \sum^n_{i=1} \left\{{\rm trace} \left(\hat V^{-1}_{\psi_j} \hat V_i\right) - {\rm trace}\left(\hat V^{-1}_{\psi_j} \hat V_i \tilde V^{-1}_i \hat V_i\right)  \right\}, \ \psi_j \in \{\tau^2,\sigma^2\}
\end{eqnarray}

\clearpage

\section*{Web Appendix B: Simulation results}
This web appendix reports a portion of the results of the simulation study carried out to evaluate the performance of Skovgaard's statistic against competing approaches, as described in Section~4 of the main manuscript.

Simulations refer to different scenarios with decreasing log event rate in the treatment group corresponding to different values for $(\beta_0, \beta_1, \mu)^\top$. Different values for the variance components $\tau^2$ and $\sigma^2$ are considered as well. The examined situations and the corresponding simulation results in terms of empirical coverage probabilities of the nominally 95\% confidence interval for $\beta_1$ are listed below.
\begin{itemize}
\item Scenario with $(\beta_0, \beta_1, \mu)^\top=(-1.5, 1, -0.5)^\top$, called \emph{scenario 2} in the main text, with $\sigma^2$ equal to 1: Figure~\ref{scenario2} \\[-5ex]
\begin{figure}[h]
\centering 
\includegraphics[width=5.5in]{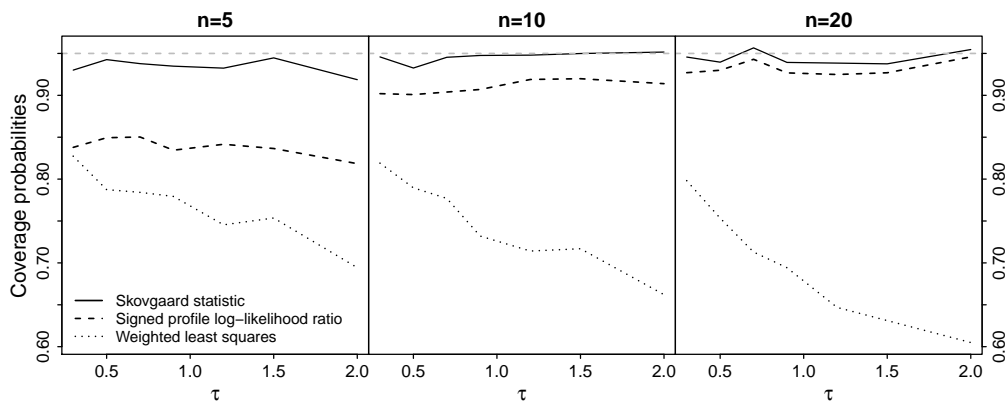} 
\caption{\small Empirical coverage probabilities of the nominally 95\% confidence interval for $\beta_1$ when $(\beta_0, \beta_1, \mu)^\top=(-1.5, 1, -0.5)^\top$, under increasing sample size $n$ and square root $\tau$ of the variance component $\tau^2$. Variance component $\sigma^2=1$. The plotted curves correspond to Skovgaard's statistic (solid), the signed profile log-likelihood ratio statistic (dashed), the weighted least squares approach (dotted). The dashed, grey horizontal line is the nominal level.}\label{scenario2}
\end{figure}
\clearpage
\newpage
\item Scenario with $(\beta_0, \beta_1, \mu)^\top=(-1.5, 1, -2.5)^\top$, called \emph{scenario 3} in the main text, with $\sigma^2$ equal to 1: Figure~\ref{scenario3} \\[-5ex]
\begin{figure}[hbtp]
\centering 
\includegraphics[width=5.5in]{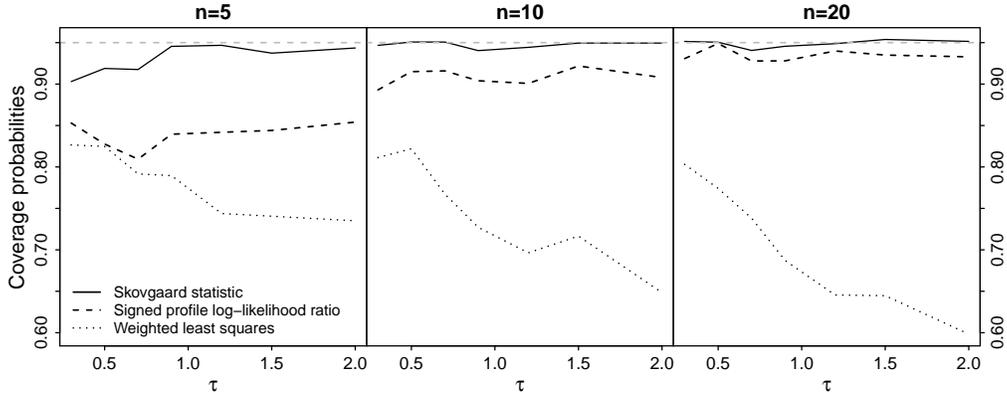} 
\caption{\small Empirical coverage probabilities of the nominally 95\% confidence interval for $\beta_1$ when $(\beta_0, \beta_1, \mu)^\top=(-1.5, 1, -2.5)^\top$, under increasing sample size $n$ and square root $\tau$ of the variance component $\tau^2$. Variance component $\sigma^2=1$. The plotted curves correspond to Skovgaard's statistic (solid), the signed profile log-likelihood ratio statistic (dashed), the weighted least squares approach (dotted). The dashed, grey horizontal line is the nominal level.}  \label{scenario3}
\end{figure}
\clearpage
\newpage
\item Scenario with $(\beta_0, \beta_1, \mu)^\top=(-3, 1, -2)^\top$, called \emph{scenario 4} in the main text, with $\sigma^2$ equal to 1: Figure~\ref{scenario4}  \\[-5ex]
\begin{figure}[hbtp]
\centering 
\includegraphics[width=5.5in]{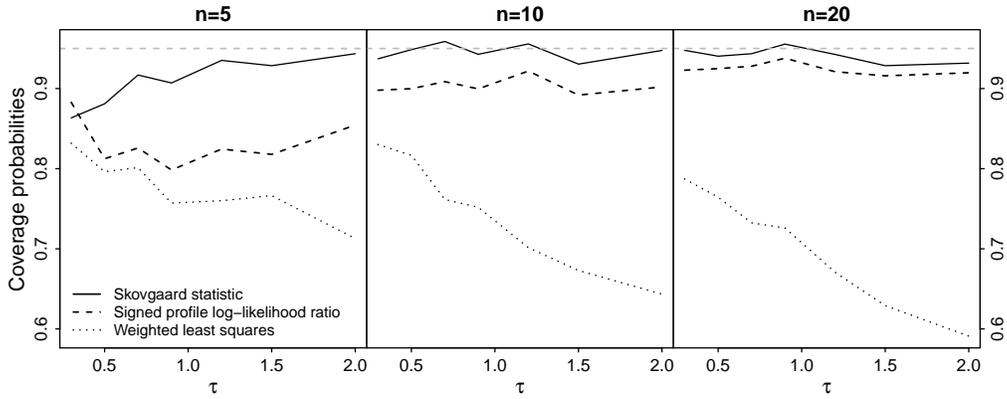} 
\caption{\small Empirical coverage probabilities of the nominally 95\% confidence interval for $\beta_1$ when $(\beta_0, \beta_1, \mu)^\top=(-3, 1, -2)^\top$, under increasing sample size $n$ and square root $\tau$ of the variance component $\tau^2$. Variance component $\sigma^2=1$. The plotted curves correspond to Skovgaard's statistic (solid), the signed profile log-likelihood ratio statistic (dashed), the weighted least squares approach (dotted). The dashed, grey horizontal line is the nominal level.}  \label{scenario4}
\end{figure}
\clearpage
\newpage
\item Scenario with $(\beta_0, \beta_1, \mu)^\top=(0, 1, 1)^\top$, called \emph{scenario 1} in the main text, with $\tau^2$ equal to 1.2: Figure~\ref{scenario1:sigma} \\[-5ex]
\begin{figure}[hbtp]
\centering 
\includegraphics[width=5.5in]{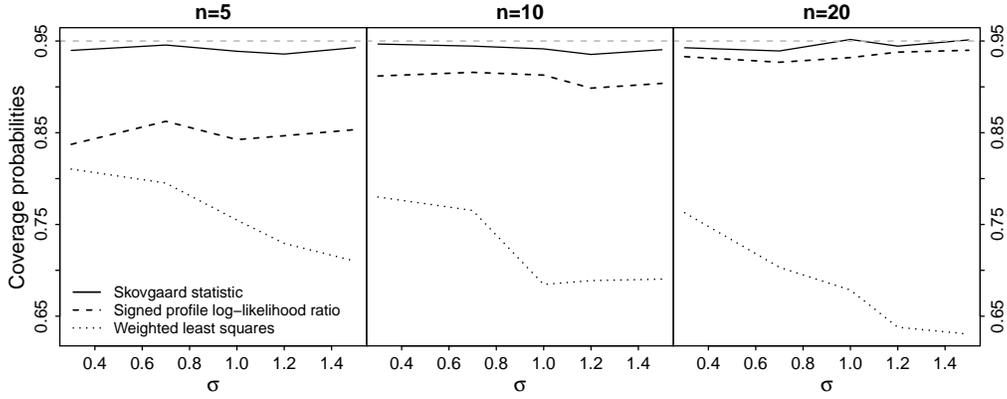} 
\caption{\small Empirical coverage probabilities of the nominally 95\% confidence interval for $\beta_1$ when $(\beta_0, \beta_1, \mu)^\top=(0, 1, 1)^\top$, under increasing sample size $n$ and square root $\sigma$ of the variance $\sigma^2$ in the control group. Variance component $\tau^2=1.2$. The plotted curves correspond to Skovgaard's statistic (solid), the signed profile log-likelihood ratio statistic (dashed), the weighted least squares approach (dotted). The dashed, grey horizontal line is the nominal level.} \label{scenario1:sigma}
\end{figure}
\clearpage
\newpage
\item Scenario with $(\beta_0, \beta_1, \mu)^\top=(-1.5, 1, -0.5)^\top$, called \emph{scenario 2} in the main text, with $\tau^2$ equal to 1.2: Figure~\ref{scenario2:sigma} \\[-5ex]
\begin{figure}[hbtp]
\centering 
\includegraphics[width=5.5in]{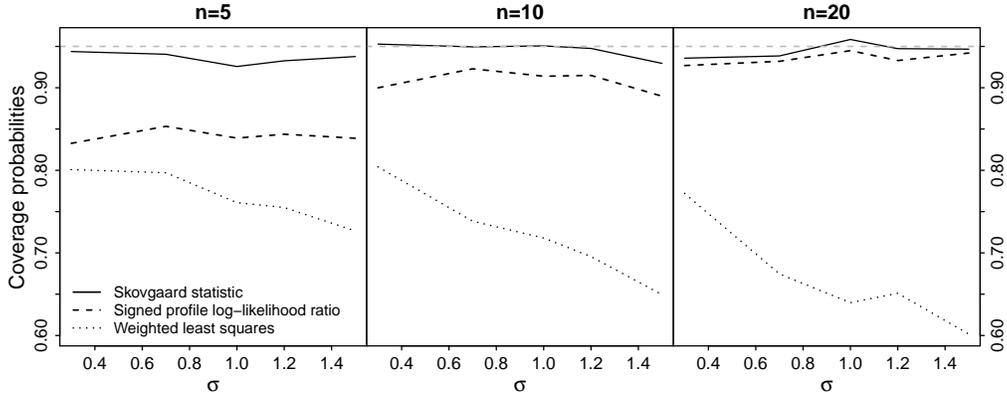} 
\caption{\small Empirical coverage probabilities of the nominally 95\% confidence interval for $\beta_1$ when $(\beta_0, \beta_1, \mu)^\top=(-1.5, 1, -0.5)^\top$, under increasing sample size $n$ and square root $\sigma$ of the variance $\sigma^2$ in the control group. Variance component $\tau^2=1.2$. The plotted curves correspond to Skovgaard's statistic (solid), the signed profile log-likelihood ratio statistic (dashed), the weighted least squares approach (dotted). The dashed, grey horizontal line is the nominal level.} \label{scenario2:sigma}
\end{figure}
\clearpage
\newpage
\item Scenario with $(\beta_0, \beta_1, \mu)^\top=(-1.5, 1, -2.5)^\top$, called \emph{scenario 3} in the main text, with $\tau^2$ equal to 1.2: Figure~\ref{scenario3:sigma} \\[-5ex]
\begin{figure}[hbtp]
\centering 
\includegraphics[width=5.5in]{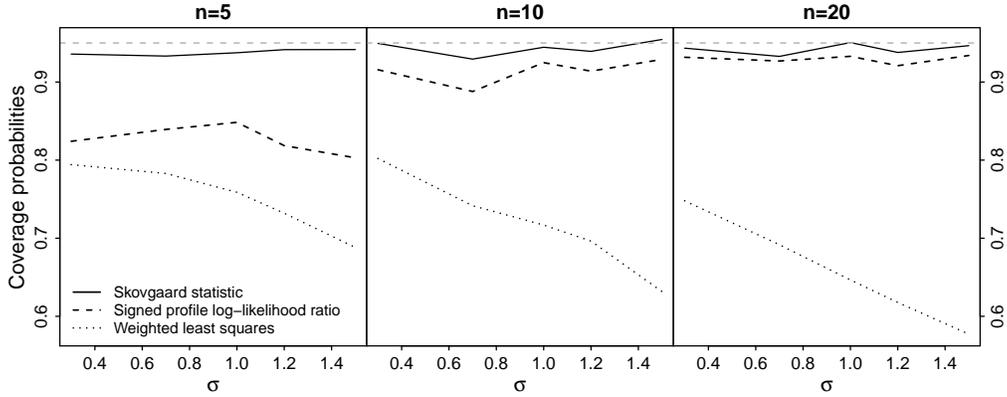} 
\caption{\small Empirical coverage probabilities of the nominally 95\% confidence interval for $\beta_1$ when $(\beta_0, \beta_1, \mu)^\top=(-1.5, 1, -2.5)^\top$, under increasing sample size $n$ and square root $\sigma$ of the variance $\sigma^2$ in the control group. Variance component $\tau^2=1.2$. The plotted curves correspond to Skovgaard's statistic (solid), the signed profile log-likelihood ratio statistic (dashed), the weighted least squares approach (dotted). The dashed, grey horizontal line is the nominal level.} \label{scenario3:sigma}
\end{figure}
\clearpage
\newpage
\item Scenario with $(\beta_0, \beta_1, \mu)^\top=(-3, 1, -2)^\top$, called \emph{scenario 4} in the main text, with $\tau^2$ equal to 1.2: Figure~\ref{scenario4:sigma}\\[-5ex]
\begin{figure}[hbtp]
\centering 
\includegraphics[width=5.5in]{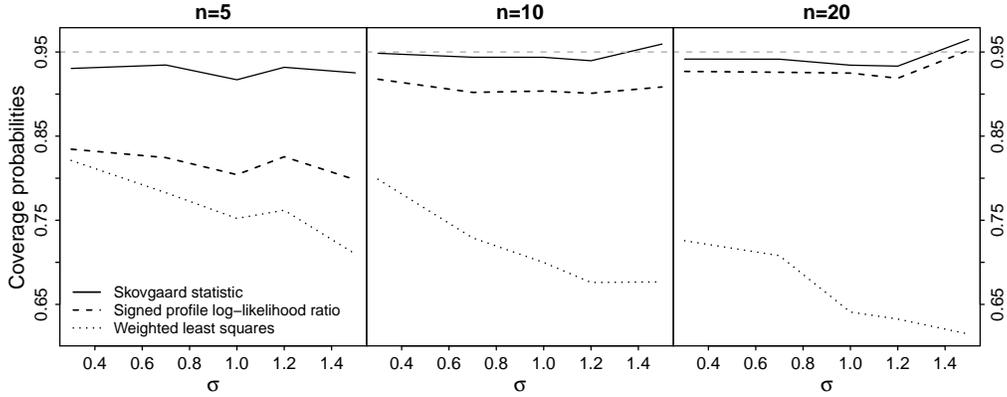} 
\caption{\small Empirical coverage probabilities of the nominally 95\% confidence interval for $\beta_1$ when $(\beta_0, \beta_1, \mu)^\top=(-3, 1, -2)^\top$, under increasing sample size $n$ and square root $\sigma$ of the variance $\sigma^2$ in the control group. Variance component $\tau^2=1.2$. The plotted curves correspond to Skovgaard's statistic (solid), the signed profile log-likelihood ratio statistic (dashed), the weighted least squares approach (dotted). The dashed, grey horizontal line is the nominal level.} \label{scenario4:sigma}
\end{figure}

\end{itemize}

\clearpage
\newpage

\section*{Web Appendix C: Data analysis}
This appendix shows how to evaluate Skovgaard's statistic for inference on the slope of the control rate regression in the \texttt R programming language. The illustration is based on the data of Hoes et al. \cite{hoes} reported in Table~1 of the paper. Functions needed to implement Skovgaard's statistic are obtained as supplementary material and they can be loaded as follows
\begin{verbatim}
R> source("control_rate_regression_LRTs.R")
\end{verbatim}
Consider the hypothesis test $\beta_1=1$ against the two-sided alternative. Wald statistic, the signed profile log-likelihood ratio statistic and Skovgaard's statistic are obtained by applying function \texttt{crr.test} (control rate regression test)
\begin{verbatim}
crr.test(data, beta1.null, alternative = c("two.sided", "less", 
    "greater"), maxit = 1000)
\end{verbatim}
with arguments
\begin{itemize}
\item \texttt{data}: the dataset
\item \texttt{beta1.null}: the value of $\beta_1$ under the null hypothesis
\item \texttt{alternative}: a character string specifying the alternative hypothesis, chosen between "two.sided" (default), "greater" or "less"; just the initial letter can be specified
\item \texttt{maxit}: the maximum number of iterations for the Nelder and Mead \cite{nelder} optimization algorithm; default value 1,000
\end{itemize}
The dataset is composed by $n$ rows corresponding to the studies recruited in the meta-analysis and $6$ columns including the values of $\hat \eta_i$, $\hat \xi_i$, and the elements of the variance/covariance matrix $\Gamma_i$ inserted by row, namely, ${\rm var}(\hat \eta_i)$, ${\rm cov}(\hat \eta_i, \hat \xi_i)$, ${\rm cov}(\hat \eta_i, \hat \xi_i)$, ${\rm var}(\hat \xi_i)$. For the analysis of Hoes et al. \cite{hoes} data, the object to be passed to function \texttt{crr.test} can be constructed as follows
\begin{verbatim}
R> deaths.treated <- c(10, 2, 54, 47, 53, 10, 25, 47, 43, 25, 157, 92)
R> ## number of person-years for the cases
R> py.treated <- c(595.2, 762, 5635, 5135, 3760, 2233, 7056.1, 8099, 
R+                 5810, 5397, 22162.7, 20885)
R> deaths.controls <- c(21, 0, 70, 63, 62, 9, 35, 31, 39, 45, 182, 72)
R> deaths.controls[2] <- 0.5
R> ## number of person-years for the controls
R> py.controls <- c(640.2, 756, 5600, 4960, 4210, 2084.5, 6824, 8267,  
R+                 5922, 5173, 22172.5, 20645)
R> py.controls[2] <- py.controls[2]+0.5
R> hoes.data.original <- data.frame(deaths.treated, py.treated,
R+                                  deaths.controls, py.controls)
## estimated log event rate for the controls
R> xi.obs <- log(hoes.data.original$deaths.treated / 
R+               hoes.data.original$py.treated)
## estimated log event rate for the treated
R> eta.obs <- log(hoes.data.original$deaths.controls / 
R+                hoes.data.original$py.controls)
R> n <- length(hoes.data.original$deaths.treated)
## variance/covariance matrix
R> gamma.matrix <- matrix(0.0, ncol=4, nrow=n) 
R> for(i in 1:n)
R+   gamma.matrix[i,] <- c(1/hoes.data.original$deaths.treated[i], 0,
R+                         0, 1/hoes.data.original$deaths.controls[i])
R> hoes.data <- data.frame(eta.obs, xi.obs, gamma.matrix)
R> colnames(hoes.data) <- c('eta.obs', 'xi.obs', 'var.eta', 'cov.etaxi',
R+                          'cov.etaxi', 'var.xi')
\end{verbatim}
Function \texttt{crr.test} 
\begin{verbatim}
R> crr.test(data=hoes.data, beta1.null=1, alternative='two.sided')

Estimate of beta1:
     Estimate  Std.Err.
WLS  0.60973   0.10892 
MLE  0.68917   0.08124 

Hypothesis test for beta1:
                                               Value       P-value   
Wald statistic                                 -3.5830787   0.0003396
Signed profile log-likelihood ratio statistic  -2.3447177   0.0190415
Skovgaard statistic                            -1.2709290   0.2037539

alternative hypothesis: parameter is different from 1
\end{verbatim}
provides the following information:
\begin{itemize}
\item the weighted least squares estimate and the maximum likelihood estimate of $\beta_1$;
\item the associated standard error;
\item the value of Wald statistic, the value of the signed profile log-likelihood ratio statistic $r_P$ and the value of Skovgaard's statistic $\overline r_P$ under the null hypothesis;
\item the p-value of the test based on the three statistics for the specified alternative hypothesis.
\end{itemize}

\newpage
\begin{verbatim}
#  Copyright 2016 Annamaria Guolo (University of Padova) 
#  Permission to use, copy, modify and distribute this software and
#  its documentation, for any purpose and without fee, is hereby granted,
#  provided that:
#  1) this copyright notice appears in all copies and
#  2) the source is acknowledged through a citation to the paper
#  Guolo A. (2016). Improving likelihood-based inference in control rate regression. Submitted.
#  The Authors make no representation about the suitability of this software
#  for any purpose.  It is provided "as is", without express or implied warranty

library(nlme)
library(mvtnorm)

## parameter vector theta:c(beta0, beta1, mu, sigma2, tau2)
## beta1.null= values of beta1 under H0
## vector of information xx = c(eta, xi, var.eta, cov.etaxi, cov.etaxi, var.xi)

crr.test <- function(data, beta1.null, alternative = c("two.sided", 
    "less", "greater"), maxit=1000){
    ans <- list()
    ans$value <- beta1.null
    alternative <- match.arg(alternative)
    ans$alternative <- alternative
    lik <- function(theta, beta1.null){
        lik.single <- function(xx, theta, beta1.null){
            p <- length(theta)
            beta0 <- theta[1]
            if(!is.null(beta1.null)) ## under H0, searching for constrained MLE
                beta1 <- beta1.null
            else
                beta1 <- theta[2]    ## searching for MLE
            mu <- theta[p-2]
            sigma2 <- theta[p-1]
            tau2 <- theta[p]
            if(any(theta[(p-1):p]<=0))
                return(NA)
            else{
                yi <- xx[1:2]
                fi <- c(beta0+beta1*mu, mu)
                Vi <- matrix(xx[3:6], ncol=2)+ matrix(c(tau2+(beta1^2)*sigma2, 
                beta1*sigma2, beta1*sigma2, sigma2), ncol=2)
                return( dmvnorm(yi, mean=fi, sigma=Vi, log=TRUE) )    
            }
        }
        values <- apply(data, 1, lik.single, theta=theta, beta1.null=beta1.null)
        return( sum(values) )
    }
    ## Mean vector for a single study
    f.single <- function(xx, theta){
        p <- length(theta)
        beta0 <- theta[1]
        beta1 <- theta[2]
        mu <- theta[3]
        sigma2 <- theta[p-1]
        tau2 <- theta[p]
        fi <- matrix(c(beta0+beta1*mu, mu), ncol=1)
        return( fi )
    }
    ## Variance/covariance matrix for a single study
    V.single <- function(xx, theta){
        p <- length(theta)
        beta0 <- theta[1]
        beta1 <- theta[2]
        mu <- theta[3]
        sigma2 <- theta[p-1]
        tau2 <- theta[p]
        Vi <- matrix(xx[3:6], ncol=2)+ matrix(c(tau2+(beta1^2)*sigma2, beta1*sigma2, 
        beta1*sigma2, sigma2), ncol=2)
        return( Vi )
    }
    ## Gradient of the mean vector for a single study
    f.grad.single <- function(xx, theta, idx){
        p <- length(theta)
        beta0 <- theta[1]
        beta1 <- theta[2]
        mu <- theta[3]
        sigma2 <- theta[p-1]
        tau2 <- theta[p]
        if(idx==1) ##beta0
            return(matrix(c(1,0), ncol=1))
        if(idx==2) ##beta1
            return(matrix(c(mu,0), ncol=1))
        if(idx==3) ##mu
            return(matrix(c(beta1,1), ncol=1))
        if(idx==4 | idx==5) ## variance components
            return( matrix(c(0,0), ncol=1) )
    }
    
    ## Gradient of the variance/covariance matrix for a single study
    V.grad.single <- function(xx, theta, idx){
        p <- length(theta)
        beta0 <- theta[1]
        beta1 <- theta[2]
        mu <- theta[3]
        sigma2 <- theta[p-1]
        tau2 <- theta[p]
        if(idx==1 | idx==3) ##beta0 o mu
            return(matrix(0.0, ncol=2, nrow=2))
        if(idx==2) ##beta1
            return(matrix(c(2*beta1*sigma2, sigma2, sigma2 ,0), ncol=2))
        if(idx==4) ##sigma2
            return(matrix(c(beta1^2, beta1, beta1, 1), ncol=2))
        if(idx==5) ##tau2
            return( matrix(c(1, 0, 0, 0), ncol=2) )
    }
    
    ## Hessian of the mean vector for a single study
    f.hess.single <- function(xx, theta, idx1, idx2){
        p <- length(theta)
        beta0 <- theta[1]
        beta1 <- theta[2]
        mu <- theta[3]
        sigma2 <- theta[p-1]
        tau2 <- theta[p]
        m <- matrix(0.0, ncol=1, nrow=2)
        if( (idx1==2 & idx2==3) | (idx1==3 & idx2==2) ) ## (beta1, mu)
            m <- matrix(c(1,0), ncol=1, nrow=2)
        return( m )
    }
    
    ## Hessian of the variance/covariance matrix for a single study
    V.hess.single <- function(xx, theta, idx1, idx2){
        p <- length(theta)
        beta0 <- theta[1]
        beta1 <- theta[2]
        mu <- theta[3]
        sigma2 <- theta[p-1]
        tau2 <- theta[p]
        m <- diag(0, 2)
        if( (idx1==2 & idx2==2) | (idx2==2 & idx1==2) ) ## (beta1, beta1)
            m <- matrix(c(2*sigma2, 0, 0, 0), ncol=2, nrow=2)
        if( (idx1==2 & idx2==4) | (idx2==2 & idx1==4) ) ## (beta1, sigma2) 
            m <- matrix(c(2*beta1, 1, 1, 0), ncol=2, nrow=2)
        return( m )
    }
    
    ## Inverse of the derivative of the variance/covariance matrix with respect to idx
    V.ginv.single <- function(xx, theta, idx){
        return( -solve(V.single(xx, theta))%*%
        V.grad.single(xx, theta, idx=idx)%*%solve(V.single(xx, theta)) )
    }
    
    ## Inverse of the Hessian of the variance/covariance matrix with respect to (idx1, idx2)
    V.hess.inv.single <- function(xx, theta, idx1, idx2){
        V <- V.single(xx, theta)
        V.idx1 <- V.grad.single(xx, theta, idx1)
        V.idx2 <- V.grad.single(xx, theta, idx2)
        V.idx1.idx2 <- V.hess.single(xx, theta, idx1, idx2)
        m <- solve(V) %*% (V.idx1%*%solve(V)%*%V.idx2 - V.idx1.idx2 + 
        V.idx2%*%solve(V)%*%V.idx1) %*% solve(V)
        return( m )
    }
    
    S.matrix <- function(theta.hat, theta.tilde){
        p <- length(theta.hat)
        S <- matrix(0.0, ncol=p, nrow=p)
        for(j in 1:2)
            for(k in 1:2)
                S[j,k] <- sum( apply(data, 1, function(x) 
                0.5*sum(diag(V.ginv.single(x, theta.hat, j)%*%
                V.single(x, theta.hat)%*%V.ginv.single(x, theta.tilde, k)%*%
                V.single(x, theta.hat))) + t(f.grad.single(x, theta.hat, j))%*%
                V.ginv.single(x, theta.tilde, k)%*%
                (f.single(x, theta.tilde)-f.single(x, theta.hat)) + 
                t(f.grad.single(x, theta.hat, j))%*%
                solve(V.single(x, theta.tilde))%*%f.grad.single(x, theta.tilde, k)) )
        for(j in 1:2)
            S[j,3] <- sum( apply(data, 1, function(x) t(f.grad.single(x, theta.hat, j))
            %*%solve(V.single(x, theta.tilde))%*%f.grad.single(x, theta.tilde, 3)) )
        for(j in 1:2)
            for(k in 4:5)
                S[j,k] <- sum( apply(data, 1, function(x) 
                0.5*sum(diag(V.ginv.single(x, theta.hat, j)%*%
                V.single(x, theta.hat)%*%V.ginv.single(x, theta.tilde, k)%*%
                V.single(x, theta.hat))) + t(f.grad.single(x, theta.hat, j))%*%
                V.ginv.single(x, theta.tilde, k)%*%
                (f.single(x, theta.tilde)-f.single(x, theta.hat))) )
        S[3,3] <- sum( apply(data, 1, function(x) t(f.grad.single(x, theta.hat, 3))%*%
        solve(V.single(x, theta.tilde))%*%f.grad.single(x, theta.tilde, 3)) )
        for(k in 4:5)
            S[3,k] <- sum( apply(data, 1, function(x) t(f.grad.single(x, theta.hat, 3))%*%
            V.ginv.single(x, theta.tilde, k)%*%
            f.single(x, theta.tilde) - t(f.single(x, theta.hat))%*%
            solve(V.single(x, theta.tilde))%*%f.grad.single(x, theta.hat, 3)) )
        for(j in 4:5)
            for(k in 4:5)
                S[j,k] <- sum( apply(data, 1, function(x) 
                0.5*sum(diag(V.ginv.single(x, theta.hat, j)%*%
                V.single(x, theta.hat)%*%V.ginv.single(x, theta.tilde, k)%*%
                V.single(x, theta.hat)))) )
        for(k in 1:2)
            S[3,k] <- sum( apply(data, 1, function(x) t(f.grad.single(x, theta.hat, 3))%*%
            V.ginv.single(x, theta.tilde, k)%*%
            (f.single(x, theta.tilde)-f.single(x, theta.hat)) + 
            t(f.grad.single(x, theta.hat, 3))%*%solve(V.single(x, theta.tilde))%*%
            f.grad.single(x, theta.tilde, k)) )
        for(j in 4:5)
            for(k in 1:2)
                S[j,k] <- sum( apply(data, 1, function(x) 
                0.5*sum(diag(V.ginv.single(x, theta.hat, j)%*%
                V.single(x, theta.hat)%*%V.ginv.single(x, theta.tilde, k)%*%
                V.single(x, theta.hat)))) )
        return(S)
    }
    
    q.vector <-  function(theta.hat, theta.tilde){
        p <- length(theta)
        q <- matrix(0.0, ncol=1, nrow=p)
        for(j in 1:p)
            q[j] <- sum(apply(data, 1, function(x) 
            0.5*sum(diag(V.ginv.single(x, theta.hat, j)%*%V.single(x, theta.hat))) - 
            0.5*sum(diag(V.ginv.single(x, theta.hat, j)%*%V.single(x, theta.hat) %*%
            solve(V.single(x, theta.tilde))%*%V.single(x, theta.hat))) + 
            t(f.grad.single(x, theta.hat, j))%*%solve(V.single(x, theta.tilde))%*%
            (f.single(x, theta.hat)-f.single(x, theta.tilde)) ))
        return(q)
    }

## expected information matrix
    i.matrix <- function(theta){
        p <- length(theta)
        i.mat <- matrix(0.0, ncol=p, nrow=p)
        for(j in 1:p)
            for(k in 1:p)
                i.mat[j,k] <- sum( apply(data, 1, function(x) 
                0.5*sum(diag(V.ginv.single(x, theta, k)%*%
                V.grad.single(x, theta, j) + solve(V.single(x, theta))%*%
                V.hess.single(x, theta, j, k) + V.hess.inv.single(x, theta, j, k)%*%
                V.single(x, theta) )) + t(f.grad.single(x, theta, j))%*%
                solve(V.single(x, theta))%*%f.grad.single(x, theta, k)) )
    return(i.mat)
            }
            
## observed information matrix
    j.matrix <- function(theta){
        p <- length(theta)
        j.mat <- matrix(0.0, ncol=p, nrow=p)
        for(j in 1:p)
            for(k in 1:p)
                j.mat[j,k] <- sum( apply(data, 1, function(x) 
                0.5*sum(diag( V.ginv.single(x, theta, k)%*%
                V.grad.single(x, theta, j) + solve(V.single(x, theta))%*%
                V.hess.single(x, theta, j, k) )) + 0.5*t(x[1:2])%*%
                V.hess.inv.single(x, theta, j, k)%*%x[1:2] - 
                t(f.hess.single(x, theta, j, k))%*%solve(V.single(x, theta))%*%x[1:2] - 
                t(f.grad.single(x, theta, j))%*%V.ginv.single(x, theta, k)%*%x[1:2] - 
                t(f.grad.single(x, theta, k))%*%V.ginv.single(x, theta, j)%*%x[1:2] - 
                t(f.single(x, theta))%*%V.hess.inv.single(x, theta, j, k)%*%x[1:2] + 
                t(f.hess.single(x, theta, j, k))%*%solve(V.single(x, theta))%*%
                f.single(x, theta) + t(f.grad.single(x, theta, j))%*%
                V.ginv.single(x, theta, k)%*%f.single(x, theta) + 
                t(f.grad.single(x, theta, j))%*%solve(V.single(x, theta))%*%
                f.grad.single(x, theta, k) + t(f.grad.single(x, theta, k))%*%
                V.ginv.single(x, theta, j)%*%f.single(x, theta) + 
                0.5*t(f.single(x, theta))%*%V.hess.inv.single(x, theta, j, k)%*%
                f.single(x, theta)) )
        return(j.mat)
    }
    
    ## correction term u
    u.stat <- function(theta.hat, theta.tilde){
        S <- S.matrix(theta.hat, theta.tilde)
        q <- q.vector(theta.hat, theta.tilde)
        j.hat <- j.matrix(theta.hat)
        j.tilde <- j.matrix(theta.tilde)
        i.hat <- i.matrix(theta.hat)
        if(det(j.hat)<0){
            j.hat <- i.hat
            print('j.hat substituted by i.hat: check the MLEs')
        }
        if(det(j.tilde[-2,-2])<0){
            j.tilde <- i.matrix(theta.tilde)
            print('j.tilde substituted by i.tilde: check the MLEs')
        }
        return( (solve(S)%*%q)[2]*sqrt( (det(j.hat)))*solve(det(i.hat))*
        det(S)*( (det(j.tilde[-2,-2])))^(-1/2) )
    }
    
    w <- 1/data$var.eta
    ## naive model, WLS
    model.naive <- lm(eta.obs~xi.obs, data=data, weights=w) ##naive model
    ## starting value for the evaluation of the MLE
    theta <- c(coef(model.naive),                 ## beta0, beta1
               mean(data[,2]),               ## mux
               var(data[,2]),        ## sigmax^2 
               (mean(resid(model.naive)^2))   ## tau^2 
               )
    ans$theta.wls <- theta
    ans$se.theta.wls <- sqrt(diag(vcov(model.naive)))
    ## Wald statistic
    wald <- (coef(model.naive)[2]-beta1.null)/sqrt(vcov(model.naive)[2,2])
    ans$wald <- wald
    ## MLE
    model.mle <- try(optim(theta, lik, control=list(fnscale=-1, maxit=maxit), 
    beta1.null=NULL), silent=TRUE)
    if(class(model.mle)=='try-error' | model.mle$convergence!=0)
        print('Possible convergence problem when searching for the MLE')
    ans$mle <- model.mle$par
    theta.hat <- model.mle$par
    se <- try(sqrt(diag(solve(i.matrix(theta.hat)))), silent=TRUE)
    ans$se.mle <- se
    model.mle.constrained <- optim(theta[-2], lik, beta1.null=beta1.null, 
    control=list(fnscale=-1, maxit=maxit))
    theta.constrained <- c( model.mle.constrained$par[1], beta1.null,  
    model.mle.constrained$par[-1])
    ## first-order statistic
    r <- sign(theta.hat[2]-beta1.null)*sqrt(2*(lik(theta.hat, beta1.null=NULL)-
    lik(theta.constrained, beta1.null=NULL)))
    u <- try(u.stat(theta.hat, theta.constrained), silent=TRUE)
    ## Skovgaard's statistic
    r.skovgaard <- r + log( (u/r) )/r
    ans$r <- r
    ans$r.skovgaard <- r.skovgaard
    if (alternative == "less") {
      ans$pvalue.wald <- pnorm(wald)
      ans$pvalue.r <- pnorm(r)
      ans$pvalue.r.skovgaard <- pnorm(r.skovgaard)
    }
    else if (alternative == "greater") {
      ans$pvalue.wald <- pnorm(wald, lower.tail = FALSE)
      ans$pvalue.r <- pnorm(r, lower.tail = FALSE)
      ans$pvalue.r.skovgaard <- pnorm(r.skovgaard, lower.tail = FALSE)
    }
    else {
      ans$pvalue.wald <- 2 * pnorm(-abs(wald))
      ans$pvalue.r <- 2 * pnorm(-abs(r))
      ans$pvalue.r.skovgaard <- 2 * pnorm(-abs(r.skovgaard))
    }
    class(ans) <- "crr.test"
    return(ans)
  }

print.crr.test <- function(x, digits = max(3L, getOption("digits") - 3L), ...){

    cat("\nEstimate of beta1:\n")
    tab <- matrix(NA, nrow=2, ncol=2)
    tab[,1] <- c(x$theta.wls[2], x$mle[2])
    tab[,2] <- c(x$se.theta.wls[2], x$se.mle[2])
    rownames(tab) <- c('WLS', 'MLE')
    colnames(tab) <- c('Estimate', 'Std.Err.')
    print.default(format(tab, digits = digits), print.gap = 2L, quote = FALSE)

    cat("\nHypothesis test for beta1:\n" )
    tab <- matrix(NA, nrow=3, ncol=2)
    tab[,1] <- c(x$wald, x$r, x$r.skovgaard)
    tab[,2] <- c(x$pvalue.wald, x$pvalue.r, x$pvalue.r.skovgaard)
    rownames(tab) <- c('Wald statistic', 'Signed profile log-likelihood ratio statistic', 
    'Skovgaard statistic')
    colnames(tab) <- c('Value','P-value')
    print.default(format(tab, digits = digits), print.gap = 2L, quote = FALSE)
    if (x$alternative == "two.sided") 
      cat("\nalternative hypothesis: parameter is different from ", 
          round(x$value, digits), sep = "", "\n")
    else cat("\nalternative hypothesis: parameter is ", x$alternative, 
             " than ", round(x$value, digits), sep = "", "\n")
  }


\end{verbatim}
\end{document}